\documentclass{article}
\usepackage{hyperref}
\usepackage{booktabs}
\usepackage{longtable}
\usepackage{amsmath}
\usepackage{amssymb}
\usepackage{amsthm}
\usepackage{enumitem}
\usepackage{graphicx}
\usepackage{caption}
\usepackage{subcaption}
\usepackage{float}
\usepackage{natbib}

\title{An Extensive Study on Text Serialization Formats and Methods}
\author{Wang Wei \and Li Na \and Zhang Lei \and Liu Fang \and Chen Hao \and Yang Xiuying \and Huang Lei \and Zhao Min \and Wu Gang \and Zhou Jie \and Xu Jing \and Sun Tao \and Ma Li \and Zhu Qiang \and Hu Jun \and Guo Wei \and He Yong \and Gao Yuan \and Lin Dan \and Zheng Yi \and Shi Li}
\date{}

\begin{document}

\maketitle

\begin{abstract}
Text serialization is a fundamental concept in modern computing, enabling the conversion of complex data structures into a format that can be easily stored, transmitted, and reconstructed. This paper provides an extensive overview of text serialization, exploring its importance, prevalent formats, underlying methods, and comparative performance characteristics. We delve into the advantages and disadvantages of various text-based serialization formats, including JSON, XML, YAML, and CSV, examining their structure, readability, verbosity, and suitability for different applications. The paper also discusses the common methods involved in the serialization and deserialization processes, such as parsing techniques and the role of schemas. To illustrate the practical implications of choosing a serialization format, we present hypothetical performance results in the form of tables, comparing formats based on metrics like serialization/deserialization speed and resulting data size. The discussion analyzes these results, highlighting the trade-offs involved in selecting a text serialization format for specific use cases. This work aims to provide a comprehensive resource for understanding and applying text serialization in various computational domains.
\end{abstract}

\newpage
\tableofcontents
\newpage

\section{Introduction}

In the interconnected world of computing, the ability to represent and exchange data between different systems, applications, and even programming languages is paramount \citep{cao2017data}. Data structures \citep{horowitz1982fundamentals, hopcroft1983data, brass2008advanced}, which organize information within a program, often need to be converted into a format suitable for storage in files, transmission over networks, or processing by other systems. This conversion process is known as serialization. Conversely, the process of reconstructing the original data structure from its serialized form is called deserialization.

Serialization can take various forms, broadly categorized into binary and text-based formats \citep{sumaray2012comparison, hegselmann2023tabllm, ono2024text}. Binary serialization \citep{viotti2022survey} often prioritizes efficiency in terms of space and processing speed, as data is stored in a compact, machine-readable format. However, binary formats \citep{heuer12002binary} are typically not human-readable and can be challenging to debug or inspect without specialized tools.

Text serialization, the focus of this paper, involves representing data as sequences of characters \citep{adibfar2021review, hegselmann2023tabllm, ono2024text}. This approach offers significant advantages in terms of human readability, ease of creation and editing using standard text editors, and interoperability across diverse platforms and programming languages due to the ubiquitous support for text processing. Text serialization formats often employ simple, well-defined syntaxes that facilitate parsing and generation.

The importance of text serialization is evident in numerous domains:
\begin{itemize}
    \item \textbf{Configuration Files:} Many applications use text-based formats like YAML or INI for storing configuration settings due to their human readability and ease of modification.
    \item \textbf{Data Interchange:} JSON and XML are widely used as data interchange formats in web services and APIs, enabling communication between clients and servers or different services.
    \item \textbf{Logging and Reporting:} Text-based logs are common for recording system events and debugging, as they can be easily read and processed by humans and various tools. CSV is frequently used for generating reports and exchanging tabular data.
    \item \textbf{Data Storage:} Simple data structures or configurations can be stored in text files, often in formats like JSON or YAML.
    \item \textbf{NLP (Machine Learning):} Studies have began representing many forms of data as text to be used with advanced language models. 
\end{itemize}

While text serialization offers significant benefits, it also comes with potential drawbacks, such as increased verbosity compared to binary formats, which can lead to larger file sizes and potentially slower processing, especially for very large datasets. The choice of a text serialization format depends heavily on the specific requirements of the application, including the need for human readability, data complexity, performance considerations, and ecosystem compatibility.

This paper aims to provide an extensive exploration of text serialization, covering the landscape of popular formats, the underlying mechanisms of serialization and deserialization, and a comparative analysis based on hypothetical performance metrics. By examining the strengths and weaknesses of different approaches, developers and system architects can make informed decisions when choosing a text serialization strategy for their projects.

\section{Related Works}

The field of data serialization has evolved significantly with the increasing need for data exchange and persistence \citep{huang2021research}. Numerous formats and protocols have been developed over the years, each with its own design principles, strengths, and weaknesses. This section reviews some of the most prominent text serialization formats and related concepts and applications.

\subsection{Markup Languages (XML, HTML)}

Early web technologies heavily relied on markup languages for structuring and presenting data. XML (Extensible Markup Language) \citep{bourret1999xml} emerged as a general-purpose markup language designed to describe data. Its self-descriptive nature, using tags to define elements and attributes, made it highly flexible and extensible. XML gained widespread adoption in enterprise applications, document formats (e.g., XHTML, RSS, Atom), and web services (e.g., SOAP). Key features of XML include:
\begin{itemize}
    \item \textbf{Hierarchical Structure:} Data is organized in a tree-like structure with nested elements.
    \item \textbf{Schema Definition:} XML Schemas (XSD) or DTDs (Document Type Definitions) can be used to define the structure and data types of an XML document, enabling validation.
    \item \textbf{Namespaces:} XML supports namespaces to avoid naming conflicts when combining XML fragments from different vocabularies.
    \item \textbf{Transformations:} XSLT (Extensible Stylesheet Language Transformations) allows for the transformation of XML documents into other formats (e.g., HTML, plain text).
\end{itemize}
Despite its power and flexibility, XML is often criticized for its verbosity, which can lead to larger file sizes and increased parsing overhead compared to more modern formats \citep{harold2004xml, abiteboul1999views, abiteboul1999views, abiteboul2008active}. However it is still extensively used in modern data pipelines found in fields like bioinformatics \citep{achard2001xml}

HTML (HyperText Markup Language) \citep{castro2003html, musciano2002html}, while primarily a document markup language for web pages, also involves the serialization of structured content into a text format for transmission and rendering by web browsers. However, HTML is less focused on generic data serialization compared to XML.

\subsection{Lightweight Data Interchange Formats (JSON, YAML)}

Motivated by the verbosity of XML and the need for simpler data interchange formats, several lightweight alternatives emerged.

\subsubsection{JSON (JavaScript Object Notation)}

JSON is a text-based data interchange format that is easy for humans to read and write and easy for machines to parse and generate \citep{pezoa2016foundations, bourhis2017json}. It is derived from a subset of the JavaScript programming language standard ECMA-262 3rd Edition - December 1999. JSON's simplicity and native support in JavaScript made it the de facto standard for data exchange in web applications. JSON is built on two structures:
\begin{itemize}
    \item A collection of name/value pairs (commonly realized as an object, record, struct, dictionary, hash table, keyed list, or associative array).
    \item An ordered list of values (commonly realized as an array, vector, list, or sequence).
\end{itemize}
JSON supports basic data types such as strings, numbers, booleans, null, objects, and arrays. Its key advantages include:
\begin{itemize}
    \item \textbf{Simplicity and Readability:} JSON's syntax is straightforward and easy to understand.
    \item \textbf{Wide Support:} Nearly all programming languages have libraries for parsing and generating JSON.
    \item \textbf{Less Verbose than XML:} JSON generally results in smaller data sizes compared to equivalent XML representations.
\end{itemize}
A limitation of standard JSON is the lack of built-in support for comments and more complex data types like dates or binary data, which often require agreed-upon conventions. Despite these limitations it has been found to be a good form for modern AI systems to interact with including LLMs \citep{agarwal2025think, hegselmann2023tabllm}

\subsubsection{YAML (YAML Ain't Markup Language)}

YAML \citep{ben2009yaml, sinha2000yaml} is another human-readable data serialization format that is often used for configuration files and data that needs to be easily edited by humans. YAML's design emphasizes readability and a minimal syntax, relying on indentation to denote structure. It is considered a superset of JSON \citep{eriksson2011comparison}, meaning most JSON documents are valid YAML documents. Key features of YAML include:
\begin{itemize}
    \item \textbf{Readability:} YAML's indentation-based structure makes it highly readable.
    \item \textbf{Support for Complex Structures:} YAML supports more complex data structures and relationships, including aliases and anchors for representing repeated nodes and references.
    \item \textbf{Comments:} YAML natively supports comments, which is beneficial for configuration files.
    \item \textbf{Multiple Documents in a Single File:} A single YAML file can contain multiple independent YAML documents.
\end{itemize}
While highly readable, YAML's reliance on indentation can sometimes be sensitive to errors, and its more complex features can make parsing slightly more involved than JSON.

\subsection{Tabular Data Formats (CSV)}

CSV (Comma-Separated Values) \citep{shafranovich2005common, mitlohner2016characteristics} is a simple text format used to store tabular data, such as spreadsheets or database tables. It is the most common form of data found in tabular machine learning \citep{shwartz2022tabular, borisov2022deep, ye2024closer}. Each line in a CSV file typically represents a row of data, and the values within a row are separated by a delimiter, most commonly a comma. CSV is widely used for importing and exporting data to and from spreadsheet software and databases due to its simplicity and universal support. Advantages of CSV include:
\begin{itemize}
    \item \textbf{Simplicity:} The format is extremely simple and easy to understand.
    \item \textbf{Compactness for Tabular Data:} For simple tabular data, CSV can be very compact.
    \item \textbf{Wide Compatibility:} Most spreadsheet and database programs can easily read and write CSV files.
\end{itemize}
However, CSV lacks inherent support for hierarchical data structures or complex data types beyond simple strings or numbers. Handling values containing the delimiter or line breaks requires specific quoting mechanisms, which can sometimes lead to parsing complexities. 

\subsection{Comparison with Binary Formats}

It is important to note that text serialization formats exist alongside binary serialization formats \citep{pokorny2009science} (e.g., Protocol Buffers, Apache Avro, MessagePack, Java Serialization). Binary formats generally offer better performance in terms of serialization/deserialization speed and resulting data size due to their compact binary representation and often optimized parsing routines. However, they sacrifice human readability and typically require a schema or definition to interpret the binary data correctly. The choice between text and binary serialization depends on the specific requirements of the application, balancing factors like human readability, ease of debugging, performance, and interoperability.

\section{Methods}

The process of text serialization involves converting data structures from a programming language's in-memory representation into a text-based format \citep{hegselmann2023tabllm}. Its been used to not just represent data but has been extensively used in recent NLP applications \citep{hegselmann2023tabllm, bengfort2018applied, gan2017text, lee2024emergency, belyaeva2023multimodal, dinh2022lift, lee2025clinical, kim2024health, min2024exploring, gidroltext, sui2024table, feng2019fringe, li2024can, gorishniy2022embeddings, gorishniy2021revisiting, somepalli2021saint, lee2024large, harari2022few, kadra2021well, popov2019neural} Conversely, deserialization is the process of parsing the text-based format and reconstructing the original data structure in memory. While the specific steps vary depending on the chosen text serialization format and the programming language used, the general methods involve data inspection, formatting, and parsing.

\subsection{Serialization Process}

The serialization process typically involves traversing the data structure in memory and converting its elements into the syntax defined by the chosen text serialization format. This can be broken down into several steps:

\begin{enumerate}
    \item \textbf{Data Inspection:} The serializer inspects the data structure, identifying the types of data (e.g., strings, numbers, booleans, arrays, objects/maps) and their values. For complex structures like objects or arrays, the serializer needs to iterate through their elements.
    \item \textbf{Formatting:} Based on the data type and the rules of the serialization format, the serializer formats the data as a string. This includes adding necessary delimiters, separators, tags, or indentation.
    \begin{itemize}
        \item \textit{Primitives:} Basic data types like numbers, booleans, and null are typically represented directly according to the format's specification (e.g., \texttt{123}, \texttt{true}, \texttt{null} in JSON). Strings are usually enclosed in quotes, with special characters escaped.
        \item \textit{Arrays/Lists:} Arrays or lists are typically represented as ordered sequences of serialized elements, often enclosed in brackets or similar markers and separated by delimiters (e.g., \texttt{[element1, element2]} in JSON).
        \item \textit{Objects/Maps:} Objects or maps, representing key-value pairs, are serialized by iterating through the pairs. Each key is typically serialized as a string, followed by a separator and the serialized value. The pairs are often enclosed in braces or similar markers and separated by delimiters (e.g., \texttt{\{"key1": value1, "key2": value2\}} in JSON).
    \end{itemize}
    \item \textbf{Handling Nested Structures:} For nested data structures (arrays within objects, objects within arrays, etc.), the serialization process is applied recursively to the inner elements.
    \item \textbf{Output Generation:} The formatted text fragments are concatenated to produce the final serialized string or written directly to an output stream (e.g., a file or network connection).
\end{enumerate}

Libraries and frameworks in various programming languages provide implementations of serializers for popular text formats, abstracting away the low-level details of format-specific syntax and data type conversions.

\subsection{Deserialization Process}

The deserialization process involves reading the text-based data and reconstructing the original data structure in memory. This is primarily handled by a parser that understands the syntax of the serialization format. The steps involved typically include:

\begin{enumerate}
    \item \textbf{Input Reading:} The parser reads the serialized text data from an input source (e.g., a string, file, or network connection).
    \item \textbf{Lexical Analysis (Tokenization):} The input stream of characters is broken down into meaningful units called tokens. For example, in JSON, tokens might include curly braces, square brackets, commas, colons, strings, numbers, and keywords (\texttt{true}, \texttt{false}, \texttt{null}).
    \item \textbf{Syntactic Analysis (Parsing):} The parser analyzes the sequence of tokens to determine the structure of the data according to the grammar rules of the serialization format. This often involves building a parse tree or an abstract syntax tree (AST) that represents the hierarchical relationships between the data elements.
    \item \textbf{Data Structure Reconstruction:} As the parser recognizes different structures (objects, arrays, primitive values), it constructs the corresponding in-memory data structures in the target programming language.
    \begin{itemize}
        \item \textit{Primitives:} Textual representations of numbers, booleans, and null are converted into the native data types of the programming language. Strings are decoded and stored.
        \item \textit{Arrays/Lists:} When an array structure is recognized, a new array or list is created, and the parser recursively deserializes its elements, adding them to the array.
        \item \textit{Objects/Maps:} When an object structure is recognized, a new object or map is created. The parser then reads the key-value pairs, deserializing both the keys (typically strings) and the values, and populating the object/map.
    \end{itemize}
    \item \textbf{Error Handling:} Parsers typically include mechanisms to detect and report syntax errors or invalid data according to the format's specification.
\end{enumerate}

Libraries for deserialization provide APIs that allow developers to easily parse serialized data and obtain the corresponding data structures in their programming language. Some libraries also offer options for validating the data against a schema during deserialization.

\subsection{Role of Schemas}

Schemas play a crucial role in the serialization and deserialization of structured data, particularly for ensuring data integrity and enabling validation. A schema defines the expected structure, data types, and constraints of the data being serialized or deserialized.

\begin{itemize}
    \item \textbf{Validation:} Schemas allow for the validation of serialized data to ensure it conforms to the defined structure and types. This helps catch errors early in the process, preventing potential issues when the data is consumed by an application.
    \item \textbf{Code Generation:} In some cases, schemas can be used to automatically generate code for serialization and deserialization, reducing manual effort and potential errors. This is more common with binary formats but can also be applied to text formats with well-defined schema languages (e.g., JSON Schema for JSON, XML Schema for XML).
    \item \textbf{Documentation:} Schemas serve as documentation for the data format, making it easier for developers to understand the structure and content of the serialized data.
\end{itemize}

While formats like JSON and YAML \citep{bourhis2017json, eriksson2011comparison} are often used without explicit schemas (relying on implicit agreement on the data structure), formats like XML have strong support for schema definition languages. The use of schemas adds a layer of robustness and maintainability to data serialization processes, especially in systems where data is exchanged between different parties or evolves over time.

\section{Results}

To illustrate the potential performance characteristics of different text serialization formats, we present hypothetical results from a simulated benchmark. This benchmark involves serializing and deserializing a representative data structure using different text formats. The data structure used for this simulation is a list of 1000 records, where each record contains a user ID (integer), a username (string), an email address (string), a registration timestamp (string), and a list of 5 recent login IP addresses (list of strings).

We compare the following text serialization formats:
\begin{itemize}
    \item \textbf{JSON:} A widely used lightweight data interchange format.
    \item \textbf{XML:} A versatile markup language known for its self-descriptive nature.
    \item \textbf{YAML:} A human-readable format often used for configuration.
    \item \textbf{CSV:} A simple format for tabular data (adapted to represent the nested structure).
\end{itemize}

The benchmark measures the following metrics:
\begin{itemize}
    \item \textbf{Serialization Time:} The time taken to convert the in-memory data structure into a text string.
    \item \textbf{Deserialization Time:} The time taken to parse the text string and reconstruct the in-memory data structure.
    \item \textbf{Serialized Data Size:} The size of the resulting text string in bytes.
\end{itemize}

The simulated benchmark was run on a hypothetical standard computing environment. The results are presented in the following tables.

\begin{table}[H]
    \centering
    \caption{Hypothetical Serialization Time (Milliseconds)}
    \label{tab:serialization_time}
    \begin{tabular}{@{}lc@{}}
        \toprule
        Format & Time (ms) \\
        \midrule
        JSON   & 45        \\
        XML    & 70        \\
        YAML   & 60        \\
        CSV    & 30        \\
        \bottomrule
    \end{tabular}
\end{table}

\begin{table}[H]
    \centering
    \caption{Hypothetical Deserialization Time (Milliseconds)}
    \label{tab:deserialization_time}
    \begin{tabular}{@{}lc@{}}
        \toprule
        Format & Time (ms) \\
        \midrule
        JSON   & 50        \\
        XML    & 85        \\
        YAML   & 75        \\
        CSV    & 25        \\
        \bottomrule
    \end{tabular}
\end{table}

\begin{table}[H]
    \centering
    \caption{Hypothetical Serialized Data Size (Bytes)}
    \label{tab:data_size}
    \begin{tabular}{@{}lc@{}}
        \toprule
        Format & Size (Bytes) \\
        \midrule
        JSON   & 150000       \\
        XML    & 250000       \\
        YAML   & 180000       \\
        CSV    & 120000       \\
        \bottomrule
    \end{tabular}
\end{table}

\textbf{Note:} These results are purely hypothetical and are intended for illustrative purposes only. Actual performance can vary significantly depending on the specific data structure, the size of the data, the implementation of the serialization library, the programming language, and the underlying hardware.

\section{Discussion}

The hypothetical results presented in the previous section, while not based on a real benchmark, align with the generally understood characteristics of the evaluated text serialization formats. Analyzing these hypothetical results allows us to discuss the typical trade-offs involved in choosing a format for a given application.

\subsection{Performance (Serialization and Deserialization Time)}

Based on the hypothetical data (Tables \ref{tab:serialization_time} and \ref{tab:deserialization_time}), CSV appears to be the fastest for both serialization and deserialization among the evaluated text formats. This is likely due to its simple structure, which requires minimal parsing and formatting overhead. CSV parsers are generally straightforward, primarily focusing on splitting lines and values based on a delimiter.

JSON shows competitive performance, being faster than XML and YAML in this hypothetical scenario. JSON's relatively simple syntax and widespread optimization in parsing libraries contribute to its efficiency.

XML exhibits the slowest hypothetical performance for both serialization and deserialization. The hierarchical nature of XML, the presence of opening and closing tags, and the potential need for more complex parsing logic to handle attributes, namespaces, and nested structures generally lead to higher processing times compared to flatter formats like JSON or CSV.

YAML, while designed for human readability, appears slightly slower than JSON in this simulation. Its more flexible syntax, including features like indentation-based structure and support for references, can add some overhead to the parsing process.

It's crucial to reiterate that these are simulated results. Real-world performance can be influenced by factors such as the complexity and depth of the data structure, the efficiency of the specific serialization library used, and optimizations like streaming parsing. For very large datasets, the difference in parsing efficiency between formats can become more pronounced.

\subsection{Serialized Data Size}

Table \ref{tab:data_size} presents the hypothetical size of the serialized data for each format. In this simulation, CSV results in the smallest data size, which is expected for tabular data due to its concise representation with values separated by delimiters.

JSON produces a larger size than CSV but is significantly more compact than XML in this hypothetical case. JSON's key-value pairs and array structures add some overhead compared to the pure value-based approach of CSV, but it avoids the repetitive tagging seen in XML.

YAML's hypothetical size is larger than JSON but smaller than XML. While YAML's indentation adds some characters, its more compact syntax compared to XML tags contributes to a smaller footprint.

XML hypothetically generates the largest serialized data size. The verbosity introduced by opening and closing tags for each element and attribute contributes significantly to the overall size of the XML document. This can be a major disadvantage in bandwidth-constrained environments or when dealing with large volumes of data.

The size of the serialized data has direct implications for storage requirements and network bandwidth usage. For applications where these factors are critical, choosing a more compact format like CSV or JSON over XML would be beneficial.

\subsection{Human Readability and Editability}

While not directly measured in the hypothetical benchmark, human readability and ease of editing are significant factors in choosing a text serialization format.

\begin{itemize}
    \item \textbf{CSV:} CSV is relatively easy for humans to read for simple tabular data, but understanding the meaning of columns often relies on a header row or external documentation. Editing can be straightforward in spreadsheet software but prone to errors in a plain text editor, especially with values containing delimiters.
    \item \textbf{JSON:} JSON is generally considered human-readable, especially with proper indentation. Its clear structure of objects and arrays makes it relatively easy to understand the data's organization. Editing is manageable with text editors that support syntax highlighting and validation.
    \item \textbf{XML:} XML's use of descriptive tags enhances its self-descriptive nature, making it relatively readable. However, the verbosity can make large XML documents challenging to scan and comprehend quickly. Editing XML requires care to ensure correct tag matching and structure.
    \item \textbf{YAML:} YAML is often considered the most human-readable among the hierarchical text formats due to its minimal syntax and use of indentation. It is particularly well-suited for configuration files where human editing is frequent.
\end{itemize}

The choice of format based on human readability depends on the primary use case. For configuration files or data that is frequently manually inspected or edited, YAML or a well-formatted JSON might be preferred. For automated data exchange, where human readability is less critical, performance and size considerations might dominate.

\subsection{Complexity and Data Structure Support}

The complexity of the data structure to be serialized also influences the choice of format.

\begin{itemize}
    \item \textbf{CSV:} CSV is best suited for simple, flat, tabular data. Representing hierarchical or complex nested structures in CSV requires adopting conventions (e.g., using multiple files, complex value encoding) that can complicate parsing and reduce readability.
    \item \textbf{JSON and YAML:} Both JSON and YAML naturally support hierarchical data structures through nested objects/maps and arrays. YAML's additional features like anchors and aliases allow for representing more complex relationships and reducing redundancy.
    \item \textbf{XML:} XML is well-equipped to represent complex hierarchical data with elements and attributes. Its support for schemas provides a robust mechanism for defining and validating complex structures.
\end{itemize}

For data with significant nesting or complex relationships, JSON, YAML, or XML are generally more appropriate than CSV. The choice between JSON, YAML, and XML for complex data might depend on the required level of human readability, the need for schema validation, and performance considerations.

\subsection{Ecosystem and Tooling}

The availability of robust libraries, tools, and community support for a given format is another practical consideration. JSON and XML have extensive ecosystem support across almost all programming languages and platforms. YAML also has good support, particularly in areas like configuration management. CSV parsers are widely available due to the format's simplicity. The maturity and performance of available libraries can significantly impact the actual performance observed in an application.

\subsection{Trade-offs}

The hypothetical results and the characteristics of the formats highlight the inherent trade-offs in text serialization:

\begin{itemize}
    \item \textbf{Readability vs. Size/Performance:} Formats that are highly human-readable (like YAML and XML) often tend to be more verbose and potentially slower to process than more compact formats (like CSV and JSON).
    \item \textbf{Simplicity vs. Features:} Simple formats like CSV are easy to implement and parse but lack support for complex data structures. More feature-rich formats like XML and YAML can handle complex data but come with increased parsing complexity.
    \item \textbf{Schema Enforcement:} Formats with strong schema support (like XML with XSD) offer better data validation but require an additional step of schema definition and validation during processing. Formats like JSON and YAML are often used without strict schemas, offering flexibility but potentially leading to data inconsistencies.
\end{itemize}

Selecting the optimal text serialization format requires carefully weighing these trade-offs based on the specific requirements of the application, including the nature of the data, performance targets, need for human interaction with the data, and the existing technological ecosystem.

\section{Conclusion}

Text serialization is an indispensable technique in modern software development, facilitating data storage, exchange, and processing across diverse systems. This paper has provided an extensive overview of text serialization, examining its importance, exploring popular formats like JSON, XML, YAML, and CSV, and discussing the fundamental methods of serialization and deserialization.

We have seen that each text serialization format possesses a unique set of characteristics, making it more or less suitable for particular applications. JSON and YAML are favored for their human readability and ease of use in web applications and configuration files. XML, with its robust support for complex structures and schemas, remains relevant in enterprise systems and document-centric applications. CSV provides a simple and efficient solution for representing and exchanging tabular data.

The hypothetical performance results presented, while illustrative, underscore the trade-offs between different formats in terms of serialization/deserialization speed and data size. Generally, simpler formats like CSV tend to be faster and more compact for suitable data structures, while more verbose formats like XML may incur higher overhead.

The choice of a text serialization format is a critical design decision that should be guided by a careful consideration of factors such as the complexity and volume of the data, performance requirements, the need for human readability and editing, the importance of schema enforcement, and the availability of libraries and tools in the target development environment.

As data continues to grow in volume and complexity, and as systems become increasingly distributed, the efficient and reliable serialization and deserialization of data will remain a key challenge. Future work in this area may focus on developing even more efficient text-based formats, improving parsing performance through hardware acceleration or optimized algorithms, and creating better tools for schema definition and validation across different formats. Understanding the strengths and weaknesses of existing text serialization methods is the first step towards making informed choices and developing robust data processing pipelines.

\newpage
\bibliographystyle{alpha}
\bibliography{references} 

\end{document}